\documentclass[pra,twocolumn]{revtex4}
\usepackage{amssymb}
\usepackage{amsfonts}
\usepackage{amsmath}
\usepackage{graphicx}
\usepackage{bm}
\usepackage{braket}

\begin{document}

\title{Solitons in Bose-Einstein Condensates with Attractive
Self-Interaction on a M\"obius Strip}
\author{Huan-Bo Luo$^{1,2}$}
\author{Guilong Li$^{1}$}
\author{Fu-Quan Dou$^{3}$}
\author{Bin Liu$^{1}$ }
\email{binliu@fosu.edu.cn}
\author{Boris A. Malomed$^{4,5}$}
\author{Josep Batle$^{6,7}$}
\author{Yongyao Li$^{1}$}
\email{yongyaoli@gmail.com}
\affiliation{$^1$School of Physics and Optoelectronic Engineering, Foshan University,
Foshan 528000, China}
\affiliation{$^2$Department of Physics, South China University of Technology, Guangzhou
510640, China}
\affiliation{$^3$College of Physics and Electronic Engineering, Northwest Normal
University, Lanzhou 730070, China}
\affiliation{$^4$Department of Physical Electronics, School of Electrical Engineering,
Faculty of Engineering, Tel Aviv University, Tel Aviv 69978, Israel}
\affiliation{$^5$Instituto de Alta Investigaci\'{o}n, Universidad de Tarapac\'{a},
Casilla 7D, Arica, Chile}
\affiliation{$^6$Departament de F\'isica UIB and Institut d'Aplicacions Computacionals de
Codi Comunitari (IAC3), Campus UIB, E-07122 Palma de Mallorca, Balearic
Islands, Spain}
\affiliation{$^7$CRISP -- Centre de Recerca Independent de sa Pobla, sa Pobla, E-07420,
Mallorca, Spain}

\begin{abstract}
We study the matter-wave solitons in Bose-Einstein condensate (BEC) trapped
on a M\"{o}bius strip (MS), based on the respective Gross-Pitaevskii (GP)
equation with the mean-field theory. In the linear regime, vortex states are
characterized by quantum numbers, $n$ and $m$, corresponding to the
transverse and circumferential directions, with the phase structure
determined by the winding number (WN) $m$. Odd and even values of $n$ must
associate, respectively, with integer and half-integer values of $m$, the
latter ones requiring two cycles of motion around MS for returning to the
initial phase. Using variational and numerical methods, we solve the GP
equation with the attractive nonlinearity, producing a family of
ground-state (GS) solitons for values of the norm below the critical one,
above which the collapse sets in. Vortex solitons with $n=1,m=1$ and $%
n=2,m=1/2$ are obtained in a numerical form. The vortex solitons with $%
n=1,m=1$ are almost uniformly distributed in the azimuthal direction, while
ones with $n=2,m=1/2$ form localized states. The Vakhitov-Kolokolov
criterion and linear-stability analysis for the GS soliton solutions and
vortices with $n=1,m=1$ demonstrates that they are completely stable, while
the localized states with $n=2,m=1/2$ are completely unstable. Finally, the
motion of solitons on the MS and the collision of two solitons are discussed.
\end{abstract}

\maketitle


\section{Introduction}

Atomic Bose-Einstein condensates (BECs) offer versatile platforms for
simulations of various phenomena from condensed-matter physics~\cite%
{hauke2012can,lewenstein2012ultracold}. Further, BECs are very accurately
modelled by Gross-Pitaevskii equations (GPEs), which are derived in the
mean-field (MF) approximation and include the inherent collisional
nonlinearity \cite{Pit,Pethick}, providing a comprehensive setting for the
studies of one-dimensional~(1D) \cite%
{achilleos2013matter,xu2013bright,salasnich2013localized,konotop2004landau,katsimiga2023solitary,kartashov2013gap,luo2022tunable}%
, two-dimensional~(2D) \cite%
{kartashov2020multidimensional,qu2016magnetic,BenLi,lobanov2014fundamental,salasnich2014localized,xu2011spin,luo2024energy,luo2022bessel,guilong2,2Ddrop,2DdropSOC}%
, and three-dimensional (3D) \cite%
{zhang2015stable,zhao2024three,malomed2022multidimensional,guilong,3Dsky}
solitons. In particular, 2D systems with the attractive local interactions
give rise to degenerate families of \textit{Townes solitons} (TSs, whose
norm does not depend on the intrinsic frequency). TSs were originally
predicted as spatial solitons in nonlinear optics~\cite{chiao1964self} and
recently observed in binary BECs~\cite%
{bakkali2021realization,chen2021observation}. This experimental result was
possible because TSs, although being unstable states, are subject to weak
(subexponential) instability against small perturbations, which may be
controlled in the experiment (although the growth of the instability
eventually leads to the critical collapse, i.e., formation of a singularity
after a finite evolution time \cite{Berge,Sulem,Fibich}).

In addition to self-trapping on the flat surface, 2D localized states can
also appear on curved ones, as recently demonstrated for solitons \cite%
{tononi2024gas}, vortices \cite{Salasnich}, and supersolids \cite{Boronat}\
on spherical surfaces. The electronic states on the pseudosphere~\cite%
{pseudosphere} have also been reported recently. The subject of the present
work is the prediction of 2D solitons -- in particular, ones with an
intrinsic topological structure -- on another well-known curved surface,
\textit{viz}., the M\"{o}bius strip (MS)~\cite{MS5,MS6}. In this connection,
it is relevant to mention that MS-shaped optical patterns (at least, as
concerns their polarization structure), that may serve as appropriate traps
for atomic BEC, have been predicted and created in a number theoretical and
experimental works \cite{MS0,MS1,MS2,MS3,MS4}. Photonic Berry-phase states
were observed in MS-shaped microcavities \cite{microcavity}. Furthermore,
BEC configurations with shapes somewhat similar to MSs were predicted and
observed in magnon condensates \cite{magnons} and atomic BECs \cite%
{BEC0,BEC1}.

MS, as a non-orientable surface with a single side and edge, has drawn
significant attention due to its unique topological properties~\cite%
{nishiguchi2018phonon,flouris2022curvature,monteiro2023dirac,liu2020quantum,huang2011topological,wang2010theoretical}%
, which preclude the application of ordinary periodic boundary conditions
(BCs) to physical models based on 2D partial differential equations. Recent
theoretical and experimental advances have revealed elementary excitations
and novel physical phenomena on MSs in a variety of settings, ranging from
molecules~\cite{walba1982total,ouyang2020self} and single crystals~\cite%
{tanda2002mobius} to optical cavities \cite{MS1,kreismann2018optical}. These
studies bring into the focus effects of the MS's local curvature effects and
its unique boundary conditions. In particular, the mean-field model for BEC
on MS, based on the respective Gross-Pitaevskii (GP) equations, offers a
natural setup for predicting solitons on curved surfaces, and may also help
to improve the understanding of the quantum mechanical behavior in 2D
systems with the MS topology.

We use mean-field theory to model the BEC on a MS in order to obtain soliton
solutions. Subsequently, in numerical calculations presented in this paper,
we handle the specific periodic BCs, introduced by MS, by dint of the
finite-element method (FEM). In the linear limit, we numerically solve the
Schr\"{o}dinger equation for MS eigenstates, including ones with embedded
vorticity. In the nonlinear regime, we obtain numerical solutions of the GP
equation the solitons (including vortical ones) and compare them to results
produced by a semi-analytical variational approximation (VA). Finally, we
develop the linear stability analysis for the solitons.

The paper is structured as follows. In Sec. II, the model and the linear
solution for the eigenstates, including vortices, are introduced. Sec. III
presents numerical soliton solutions of the GP equation with the attractive
interaction, representing the MS ground state (GS), along with approximate
soliton solutions obtained through VA, which are found to closely match the
numerical counterparts. Sec. IV reports the numerically found vortex soliton
solutions of the GP equation. The stability of the GS and vortex solitons
against small perturbations is discussed in Sec. V. In Sec. VI, the motion
of solitons on the MS and the collision of two solitons are discussed.
Finally, Sec. VII concludes the paper.

\section{The model and solutions of the linear system}

We consider atomic BEC with attractive interactions, loaded onto MS with
width $2w$ and radius $r$, as shown below in Figs. \ref{figure1}-\ref%
{figure3}. By means of scaling, we fix $r=2$ and then focus of the value of $%
w=1$, . The Cartesian coordinates on MS are written in a parametric form as
\begin{align}
x(u,v)& =\left( r+wv\cos \frac{u}{2}\right) \cos u,  \notag \\
y(u,v)& =\left( r+wv\cos \frac{u}{2}\right) \sin u,  \label{xyz} \\
z(u,v)& =wv\sin \frac{u}{2},  \notag
\end{align}%
where curvilinear coordinates $\ u$ and $v$ take values in intervals $-\pi
\leq u\leq \pi $ and $-1\leq v\leq 1$. This parametrization defines MS
centered at the origin of the $(x,y)$ plane.

In terms of coordinates (\ref{xyz}), the MS metric $g$ and Jacobian
determinant $|J|$ are given by
\begin{equation}
g=%
\begin{pmatrix}
\lambda ^{2} & 0 \\
0 & w^{2}%
\end{pmatrix}%
,~|J|=\sqrt{|g|}=w\lambda ,  \label{metric}
\end{equation}%
where
\begin{equation}
\lambda \left( u,v\right) \equiv \sqrt{\left( r+wv\cos \frac{u}{2}\right)
^{2}+\frac{w^{2}v^{2}}{4}}.  \label{lambda}
\end{equation}%
Accordingly, the free-space GP equation, $i\partial \Psi /\partial
t=-(1/2)\nabla ^{2}\Psi -|\Psi |^{2}\Psi $, constrained onto the MS, takes
the form of
\begin{equation}
i\frac{\partial \Psi }{\partial t}=-\frac{1}{2\lambda }\partial _{u}\left(
\frac{1}{\lambda }\partial _{u}\Psi \right) -\frac{1}{2w^{2}\lambda }%
\partial _{v}\left( \lambda \partial _{v}\Psi \right) -|\Psi |^{2}\Psi .
\label{main}
\end{equation}

Applying rescaling to Eqs. (\ref{main}) and (\ref{lambda}), \textit{viz}.,
\begin{equation}
w\equiv (r/2)\tilde{w},\lambda \equiv (r/2)\tilde{\lambda},t\equiv (r/2)^{2}%
\tilde{t},\Psi \equiv (2/r)\tilde{\Psi},  \label{scaling}
\end{equation}
these equations are reduced to ones with $r=2$, which is fixed below. Then,
we focus on the case of $w=1$, which makes it possible to produce generic
results, and drop tildes over the rescaled variables.

Stationary solutions of Eq.~\eqref{main} with real chemical potential $\mu $
are sought for in the usual form,
\begin{equation}
\Psi (u,v,t)=\exp (-i\mu t)\psi (u,v),  \label{psipsi}
\end{equation}%
with function $\psi (u,v)$ satisfying equation
\begin{equation}
\mu \psi =-\frac{1}{2\lambda }\partial _{u}\left( \frac{1}{\lambda }\partial
_{u}\psi \right) -\frac{1}{2w^{2}\lambda }\partial _{v}\left( \lambda
\partial _{v}\psi \right) -|\psi |^{2}\psi .  \label{stationary}
\end{equation}%
In terms of coordinates (\ref{xyz}) and metric coefficient (\ref{lambda}),
the norm (scaled particle number) and energy (Hamiltonian) of the stationary
states are defined as%
\begin{equation}
N=\int_{-\pi }^{\pi }du\int_{-1}^{1}dv\left\vert \psi \right\vert
^{2}w\lambda ,  \label{N}
\end{equation}%
\begin{equation}
E=\int_{-\pi }^{\pi }du\int_{-1}^{1}dv\left( \frac{w}{2\lambda }|\partial
_{u}\psi |^{2}+\frac{\lambda }{2w}|\partial _{v}\psi |^{2}-\frac{w\lambda }{2%
}|\psi |^{4}\right) .  \label{E}
\end{equation}%
The stationary wave function $\psi (u,v)$ needs to satisfy the natural BCc
for MS, which are
\begin{equation}
\begin{array}{c}
\psi (u,v=-1)=\psi (u,v=1)=0, \\
\psi (u=-\pi ,v)=\psi (u=\pi ,-v),%
\end{array}
\label{BC}
\end{equation}%
that can be easily handled by FEM, see Appendix A for details.

\begin{figure}[tbp]
\centering
\includegraphics[width=3.4in]{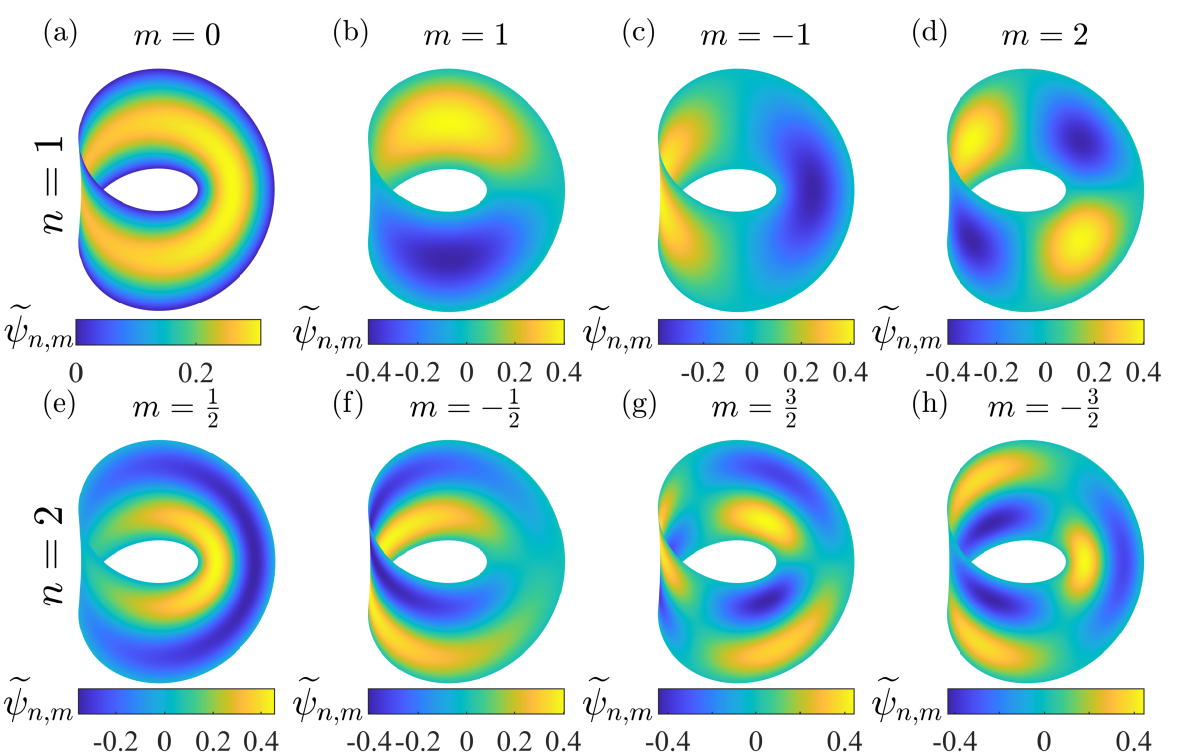}
\caption{Real MS eigenstates $\widetilde{\protect\psi }_{n,m}$, as produed
by the linearized equation \eqref{stationary} in the form of the angular
standing waves, which are defined as per Eq. (\protect\ref{eigenstate}). The
eigenstates are plotted for the lowest values of the transverse and magnetic
quantum numbers, $n$ and $m$, which are indicated in panels.}
\label{figure1}
\end{figure}

We first address the linearized version of Eq.~\eqref{stationary}, looking
for vortex states as $\psi _{n,m}(u,v)$, where $n=1,2,3,\ldots $ is the
quantum number in the transverse ($v$) direction, and
\begin{equation}
m=\left\{
\begin{array}{ll}
0,\pm 1,\pm 2,\pm 3,\ldots & \text{if }n\text{ is odd,} \\
\pm \frac{1}{2},\pm \frac{3}{2},\pm \frac{5}{2},\ldots & \text{if }n\text{
is even,}%
\end{array}%
\right.  \label{m}
\end{equation}%
is the winding number (WN, alias vorticity, or magnetic quantum number) in
the circumferential ($u$) direction. The WN sign indicates if the phase of
the wave function increases or decreases as it rotates counterclockwise
along the $u$ direction. Additionally, the vortex state satisfies constraint
\begin{equation}
\psi _{n,m}^{\ast }(u,v)=\psi _{n,-m}(u,v),  \label{conj}
\end{equation}%
with $\ast $ standing for the complex conjugate. The energy spectrum
produced by this linear Schr\"{o}dinger equation was found in a numerical
form in Ref. \cite{li2012quantum}, where it was demonstrated that it may be
accurately approximated by formula
\begin{equation}
\mu _{n,m}=\frac{\pi ^{2}n^{2}}{8w^{2}}+\frac{m^{2}}{2r^{2}}.
\label{spectrum}
\end{equation}%
The spectrum is subject to the exact relation, $\mu _{n,m}=\mu _{n,-m}$,
which means that the system is always doubly degenerate, except for $m=0$.
The degeneracy implies that the linear system admits, instead of the set of
vortical eigenmodes $\psi _{n,\pm m}$, which are similar to $\exp \left( \pm
imu\right) $, an alternative set in the form of angular standing waves,
which are similar to $\cos \left( mu\right) $ and $\sin \left( mu\right) $,
\textit{viz}.,
\begin{equation}
\widetilde{\psi }_{n,m}=\left\{
\begin{array}{ll}
\left( \psi _{n,-m}+\psi _{n,m}\right) /\sqrt{2}, & \text{if }m>0, \\
\psi _{n,m}, & \text{if }m=0, \\
i\left( \psi _{n,m}-\psi _{n,-m}\right) /\sqrt{2}, & \text{if }m<0.%
\end{array}%
\right.  \label{eigenstate}
\end{equation}%
It follows from Eqs.~\eqref{conj} and ~\eqref{eigenstate} that wave
functions (\ref{eigenstate}) obey the construct $\widetilde{\psi }%
_{n,m}^{\ast }=\widetilde{\psi }_{n,m}$, hence they are real-valued
functions. The lowest numerically found eigenstates $\widetilde{\psi }_{n,m}$
are shown in Fig.~\ref{figure1}, for $n=1$ and $n=2$. As these eigenstates $%
\widetilde{\psi }_{n,m}$ were studied in detail in Ref.~\cite{li2012quantum}%
, we do not elaborate on them here.

\begin{figure}[tbp]
\centering
\includegraphics[width=3.4in]{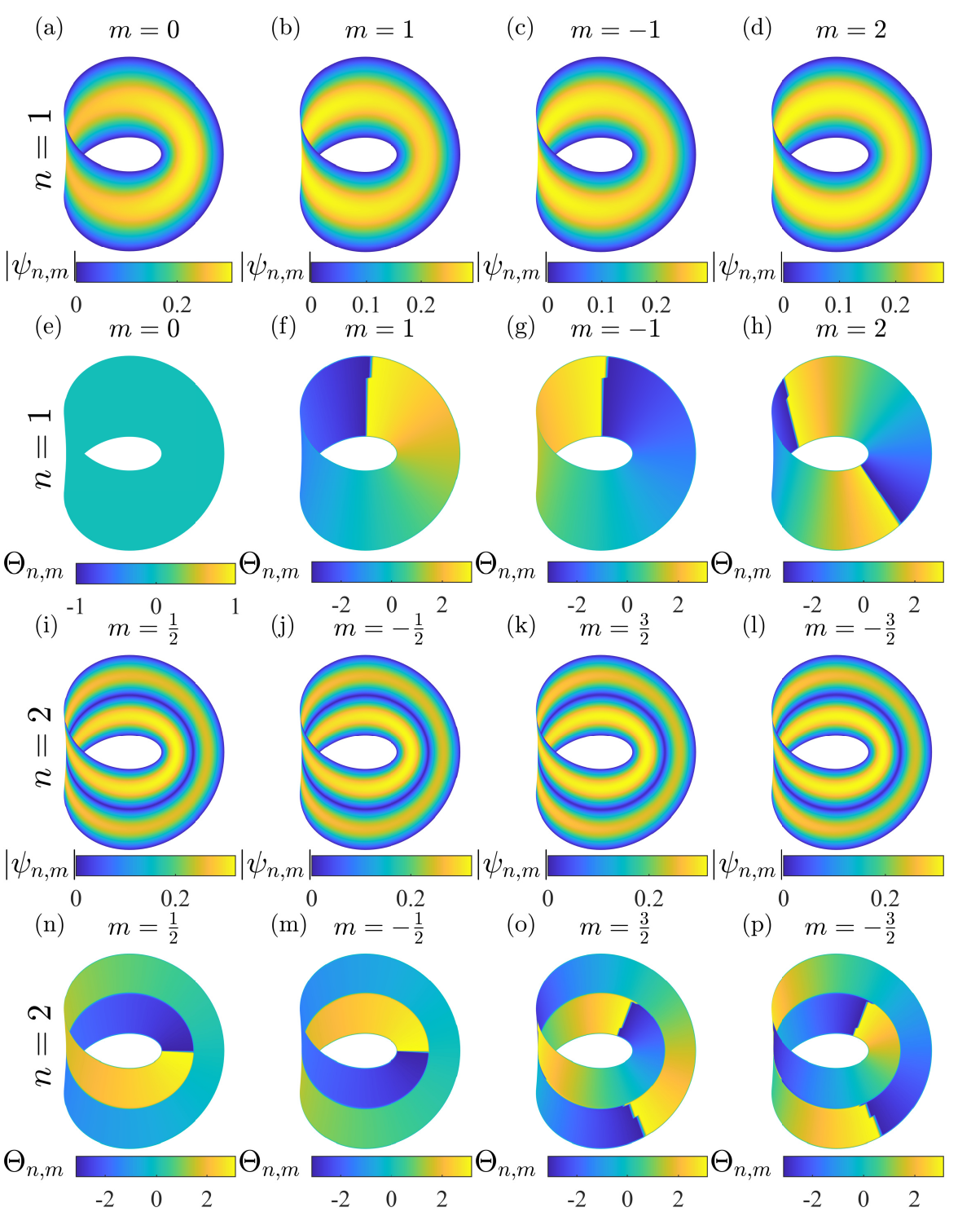}
\caption{The density and phase distributions of complex MS vortex states $%
\protect\psi _{n,m}$, as produced by the linearized equation
\eqref{stationary}. The eigenstates are plotted for values of the transverse
and magnetic quantum numbers, $n$ and $m$, which are indicated in panels.}
\label{figure2}
\end{figure}

Inverting Eq.~\eqref{eigenstate}, one can construct the linear vortex
eigenstates $\psi _{n,m}$ from those $\widetilde{\psi }_{n,m}$. They are
displayed in Fig.~\ref{figure2}, in terms of $|\psi _{n,m}|$ and their phase
$\Theta _{n,m}$. It is seen that their density distributions are almost
identical for the vortex states with the same $n$. For odd values of $n$,
such as $n=1$ in Fig.~\ref{figure2}, WN $m$ is an integer, as in traditional
vortex states. This means that, walking once around MS. one will retrieve
the initial phase. However, for even values of $n$, such as $n=2$ in Fig.~%
\ref{figure2}, $m$ is half-integer, implying that one needs to travel twice
around MS for the retrieval of the initial phase.

\section{The numerical solution and VA (variational approximation) for the
GS solitons of the nonlinear system}

Next, we address the complete form of Eq.~\eqref{stationary}, including the
nonlinear term. The use of FEM makes it possible to satisfy BCs \eqref{BC}
in this case too. Here, in the framework of the FEM technique, we use a very
efficient Newton's algorithm to obtain the ground-state (GS) solution in the
case of the self-attractive nonlinearity in Eq. (\ref{stationary}).

To apply the Newton's algorithm, we use the above-mentioned linear
eigenstate $\psi _{1,0}$, with the corresponding approximate linear
eigenvalue $\mu _{1,0}=1.23$, as given by Eq. (\ref{spectrum}) (recall we
set $w=1$), as the initial guess. Then, the Newton's iterations quickly
converge to a numerically accurate solution for the wave function.

\begin{figure}[tbp]
\centering
\includegraphics[width=3.4in]{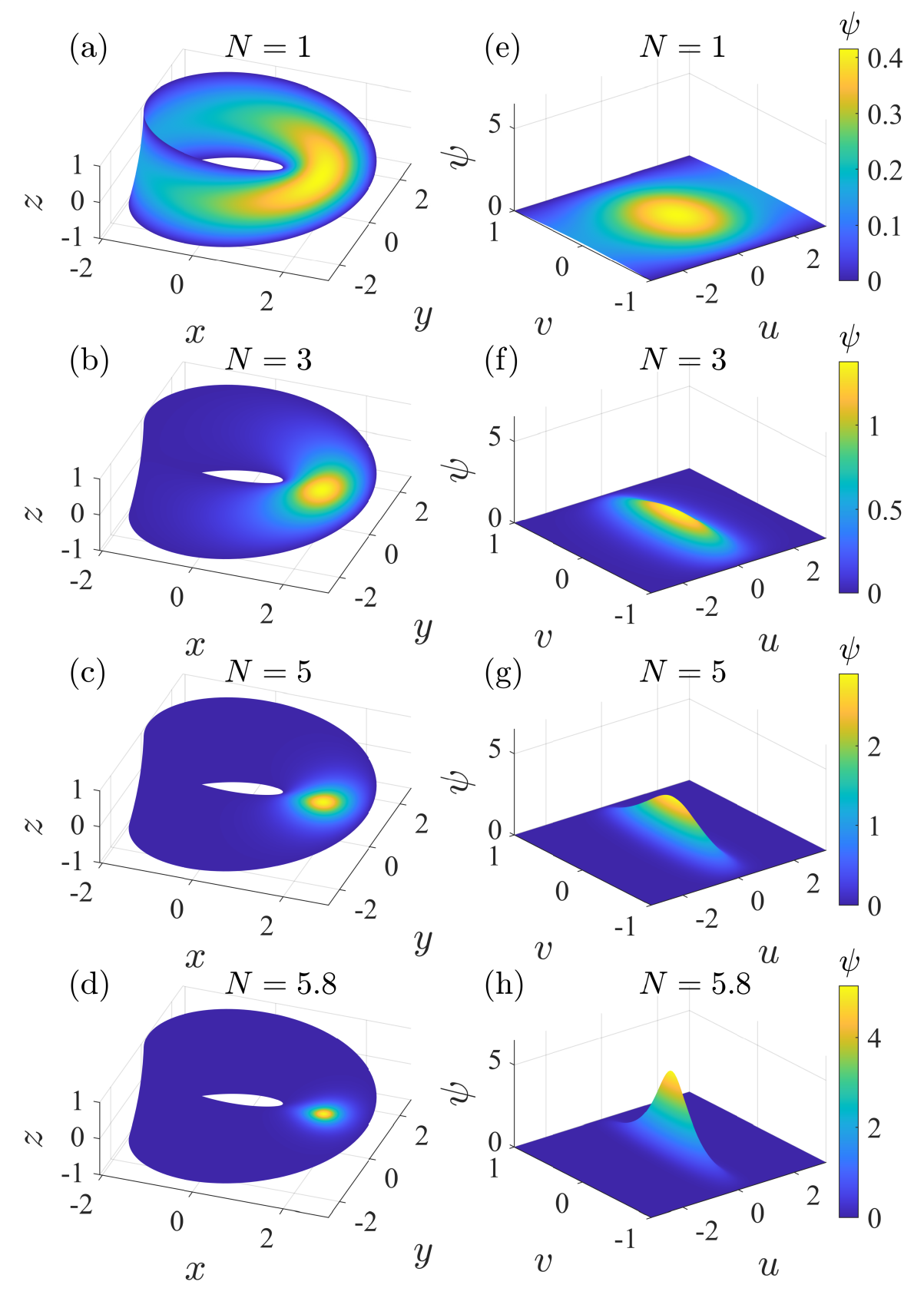}
\caption{Real GS solutions of the full nonlinear equation (\protect\ref%
{stationary}), corresponding to quantum numbers ($n=1,m=0$), with norms and
chemical potential (a, e) $N=1,~\protect\mu =1.16$, (b, f) $N=3,~\protect\mu %
=0.56$, (c, g) $N=5,~\protect\mu =-1.31$ and (d, h) $N=5.8,~\protect\mu %
=-5.40$.}
\label{figure3}
\end{figure}

By gradually decreasing the chemical potential $\mu $, the real GS wave
functions for different values of norm $N$ (\ref{N}) are obtained and
plotted in the Cartesian coordinates $\left( x,y,z\right) $ and MS ones $%
\left( u,v\right) $ [see Eq. (\ref{xyz})] in Figs.~\ref{figure3}(a-d) and
(e-h), respectively. In the $\left( x,y,z\right) $ space, the size of the
self-trapping area decreases with the increase of $N$, due to the
strengthening self-attraction. Simultaneously, in the $\left( u,v\right) $
space, the wave-function's peaks become increasingly sharper with the
increase of $N$.

The self-trapping can be roughly determined by analyzing the
interaction-energy term in expression (\ref{E}):
\begin{equation}
E_{\mathrm{int}}=-\frac{w}{2}\int_{-\pi }^{\pi }du\int_{-1}^{1}dv\left[
\lambda \left( u,v\right) |\psi |^{4}\right] .  \label{EI}
\end{equation}%
To minimize $E_{\mathrm{int}}$, density $|\psi |^{2}$ needs to concentrate
at the location where $\lambda (u,v)$ has its maximum, i.e., around $u=0$
and $v=1$, as per Eq. (\ref{lambda}). However, due to BC~\eqref{BC}, which
dictates $\psi (u,v=1)=0$, the density concentrates around $u=0$, avoiding $%
v=1$. In the Cartesian coordinates, $u=0$ corresponds to $x=r+wv$, $y=0$,
and $z=0$, according to Eq. (\ref{xyz}). These expectations agree with the
numerical findings presented in Fig. \ref{figure3}.

The family of the GS solutions is naturally characterized by the dependence
of its norm $N$ on chemical potential $\mu $, which is plotted in Fig. \ref%
{figure4}(a). Under the action of the self-attraction, $N$ initially rapidly
grows as $\mu $ decreases (towards $\mu <0$). Then, the growth saturates,
and $N$ attains the limit value for the extremely narrow soliton, $N_{%
\mathrm{TS}}(\mu =-\infty )\approx 5.85$, which is the commonly known value
of the norm of the above-mentioned TSs in the flat 2D space \cite%
{chiao1964self,Berge,Fibich}. Indeed, for the solitons with the vanishingly
small size the difference between the flat space and MS is negligible. In
this connection, it is relevant to mention that the norm of the TSs with
embedded integer vorticity $m$, found in the 2D flat space \cite%
{Minsk1,Minsk2}, increases with the growth of $m$, so that $N_{\mathrm{TS}%
}^{(m=1)}\approx 24.1$ and $N_{\mathrm{TS}}^{(m=2)}\approx 44.8,$ in
agreement with the approximate analytical prediction \cite{ECNU},%
\begin{equation}
\left( N_{\mathrm{TS}}^{(m\geq 1)}\right) _{\mathrm{analyt}}\approx 4\sqrt{3}%
\pi m.  \label{analyt}
\end{equation}

\begin{figure}[tbp]
\centering
\includegraphics[width=3.4in]{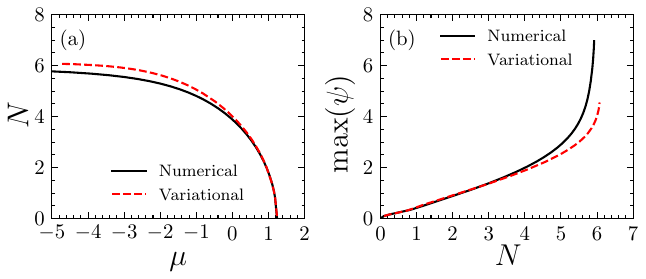}
\caption{(a) The norm of the GS solitons obtained from numerical and
variational methods vs. the chemical potential $\protect\mu $. (b) The
amplitude of the GS solitons, obtained in the numerical and variational
forms, $\max (\protect\psi )$, vs. the norm.}
\label{figure4}
\end{figure}

The fact that the plot of $N(\mu )$ in Fig. \ref{figure4}(a) obeys the
\textit{Vakhtiov-Kolokolov (VK)\ criterion}, $dN/d\mu <0$, implies the GS
soliton family satisfies the corresponding necessary stability condition
\cite{VakhKol,Berge,Sulem,Fibich} (the TS family in the flat 2D space is
degenerate, with $dN/d\mu =0$, hence they do not obey the VK criterion, and
are (weakly) unstable, as mentioned above). The full stability analysis for
the GS solitons is reported below in Sec. VII.

A related dependence of the GS-soliton's amplitude, $\text{max}(\psi )$, on $%
N$ is plotted in Fig.~\ref{figure4}(b). The divergence of the amplitude at $%
N\rightarrow N_{\mathrm{TS}}$ corroborates the above-mentioned fact that the
GS becomes infinitely narrow in this limit.

In addition to the numerical GS solution presented above, it is possible to
apply VA with the aim to approximate the solution by a relevant analytical
expression. To this end, the variational \textit{ansatz} is chosen in the
following tractable form:
\begin{equation}
\psi _{\mathrm{VA}}(u,v)=A(1-v^{2})\mathrm{sech}(au)\exp \left[
-b(v-v_{0})^{2}\right],  \label{ansatz}
\end{equation}%
with real parameters $A$, $a$, $b>0$, and $v_{0}$ representing the soliton's
amplitude, its inverse widths in the $u$ and $v$ directions, respectively,
and the offset in the $v$ direction. Factor $(1-v^{2})$ is introduced in the
ansatz to secure that it satisfies the first BC (\ref{BC}). Strictly
speaking, the ansatz does not satisfy the second BC from Eq. (\ref{BC});
further analysis demonstrates that this BC holds approximately in the case
of $N>2$, in which case the soliton is narrow enough to produce negligibly
small values of $\psi (\pm \pi ,v)$.

Because VA is developed for a given value of $N$, amplitude $A$ is not an
independent parameter, being expressed in terms of $a$, $b$, and $v_{0}$ as
per Eq.~\eqref{N}, with $\psi $ substituted by ansatz (\ref{ansatz}). The
elimination of $A$ was performed by means of numerical computation of the
norm integral (\ref{N}), as it cannot be performed in an analytical form.

The substitution of the VA ansatz~\eqref{ansatz} in expression~\eqref{E}
produces the GS-soliton's energy as a function of the variational parameters
$a$, $b$ and $v_{0}$:
\begin{gather}
E_{\mathrm{VA}}(a,b,v_{0})=\int_{-\pi }^{+\pi }du\int_{-1}^{+1}dv\left\{
\frac{wa^{2}\mathrm{tanh}^{2}(au)}{2\lambda }\psi _{\mathrm{VA}}^{2}\right.
\notag \\
\left. \quad +\frac{2}{w}\lambda \left[ \frac{v}{1-v^{2}}+b(v-v_{0})\right]
^{2}\psi _{\mathrm{VA}}^{2}-\frac{w}{2}\lambda \psi _{\mathrm{VA}%
}^{4}\right\} .  \label{Eab}
\end{gather}%
In the framework of VA, values of the parameters for the GS soliton with
fixed norm $N$ are predicted as those at which energy $E_{\mathrm{VA}}$,
given by Eq. (\ref{Eab}), attains its minimum [recall parameter $A$ of the
VA ansatz (\ref{ansatz}) is eliminated in favor of $N$]. Under the latter
condition, the minimum of expression (\ref{Eab}) was found numerically, by
means of the simplex method \cite{simplex,simplex2}.

\begin{figure}[tbp]
\centering
\includegraphics[width=3.4in]{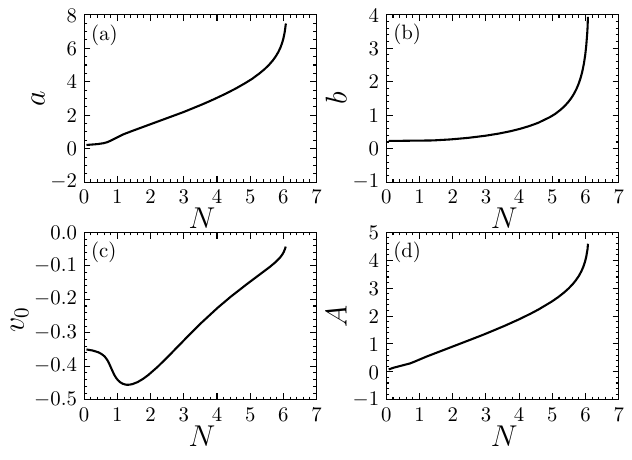}
\caption{Parameters of ansatz (\protect\ref{ansatz}) vs. the norm, as
predicted by VA: (a,b) the inverse widths $a$ and $b$ in the $u$- and $v$%
--directions; (c) offset $v_{0}$ in the $v$-direction; (d) normalization
factor $A$. }
\label{figure5}
\end{figure}

\begin{figure}[bp]
\centering
\includegraphics[width=3.4in]{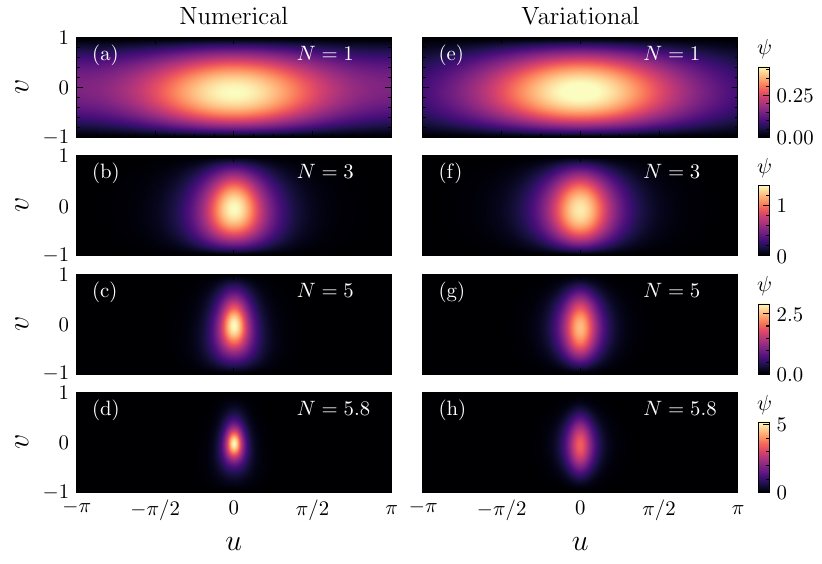}
\caption{Comparison of the VA-predicted shapes of the GS solitons at
different values of norm $N$ with the numerically found solitons, for the
same values of $N$.}
\label{figure6}
\end{figure}

Dependences of the VA-predicted parameters in ansatz (\ref{ansatz}) on $N$
are plotted in Fig.~\ref{figure5}, and the respective dependences $N(\mu )$
and $\max (\psi )$ vs. $N$ are shown in Fig. \ref{figure4}. As illustrated
in Fig. \ref{figure4}, the curves produced by the VA are very close to their
counterparts provided by the numerical results. In particular, they exhibit
full overlapping at $N<4$. Naturally, the GS soliton becomes narrower in the
$u$ and $v$ directions with the increase of $N$, while its amplitude
increases. The VA solution for the GS soliton does not exists at $%
N>6.06\equiv \left( N_{\mathrm{TS}}\right) _{\mathrm{VA}}$, which is the VA
counterpart of the above-mentioned limit value, $N_{\mathrm{TS}}\approx 5.85$
[the simple VA for the TSs in the flat 2D space yields a coarser VA
prediction, $\left( N_{\mathrm{TS}}\right) _{\mathrm{VA}}=2\pi $ \cite%
{Gothenburg}, while for the TSs with embedded integer vorticity $m\geq 1$ an
analytical approximation is provided by the above-mentioned expression (\ref%
{analyt})].

We compare the VA-predicted results to their numerically found counterparts
in Fig.~\ref{figure6} for norms $N=1,~3,~5,$ and $5.8$ (the latter value is
taken close to the critical one, $N_{\mathrm{TS}}\approx 5.85$, above which
the GS solitons do not exist, as said above). The corresponding values of
the variational parameters are presented in Table~\ref{parameters}. We
conclude that the VA-predicted shape of the GS wave function is very close
to its numerical counterparts for $N<4$, while for $N\geq 4$ the numerically
found wave function exhibits stronger localization and a sharper peak. This
discrepancy is explained by the fact that the ansatz (\ref{ansatz}) is not
sufficiently accurate for very narrow solitons.

\begin{figure}[tbp]
\centering
\includegraphics[width=3.4in]{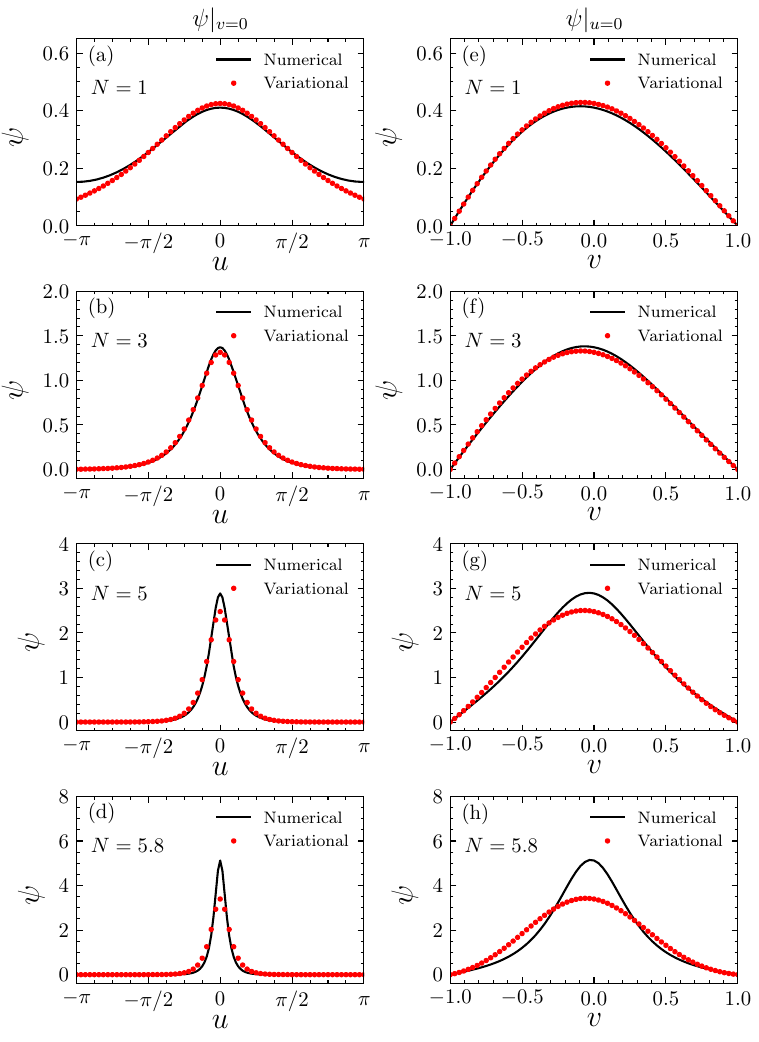}
\caption{The comparison of cross sections (along $v=0$ and $u=0$ in the left
and right columns, respectively) of the variational and numerical GS soliton
solutions, for different values of norm $N$.}
\label{figure7}
\end{figure}

\begin{table}[h]
\centering
\renewcommand{\arraystretch}{1.5}
\par
\begin{tabular}{@{\hspace{12pt}}c@{\hspace{24pt}}c@{\hspace{24pt}}c@{\hspace{24pt}}c@{\hspace{24pt}}c}
\hline
$N$ & $a$ & $b$ & $v_0$ & $A$ \\ \hline
1 & 0.701 & 0.234 & $-0.440$ & 0.445 \\
3 & 2.195 & 0.389 & $-0.323$ & 1.370 \\
5 & 4.104 & 0.987 & $-0.147$ & 2.533 \\
5.8 & 5.611 & 1.962 & $-0.085$ & 3.445 \\ \hline
\end{tabular}%
\caption{Values of parameters ($a,b,v_{0},A$) in ansatz (\protect\ref{ansatz}%
), as produced by VA for values of norm $N$ presented in Fig. \protect\ref%
{figure6}.}
\label{parameters}
\end{table}

More detailed comparison of the VA-predicted shapes of the GS solitons and
their numerical counterparts is presented in Fig. \ref{figure7} by means of
cross-sections of the solitons along $v=0$ and $u=0$. For small values of
the norm, such as $N=1$, the discrepancy between the variational and
numerical results mainly occurs near $u=\pm \pi $, because, as mentioned
above, ansatz~\eqref{ansatz} does not accurately satisfy the second BC
in~Eq. \eqref{BC} at $N<2$. At intermediate values of the norm, such as $N=3$%
, VA agrees well with the numerical solution. Finally, as is also mentioned
above, at values of $N$ close to the existence threshold ($N=N_{\mathrm{TS}}$%
) , such as $N=5$ and $N=5.8$, the numerical solution exhibits a sharper
peak in comparison to the VA prediction, as ansatz (\ref{ansatz}) is not
quite accurate in this case. A general conclusion is that the relatively
simple version of VA, based on ansatz (\ref{ansatz}), produces quite
reasonable results.

\begin{figure}[tbp]
\centering
\includegraphics[width=3.4in]{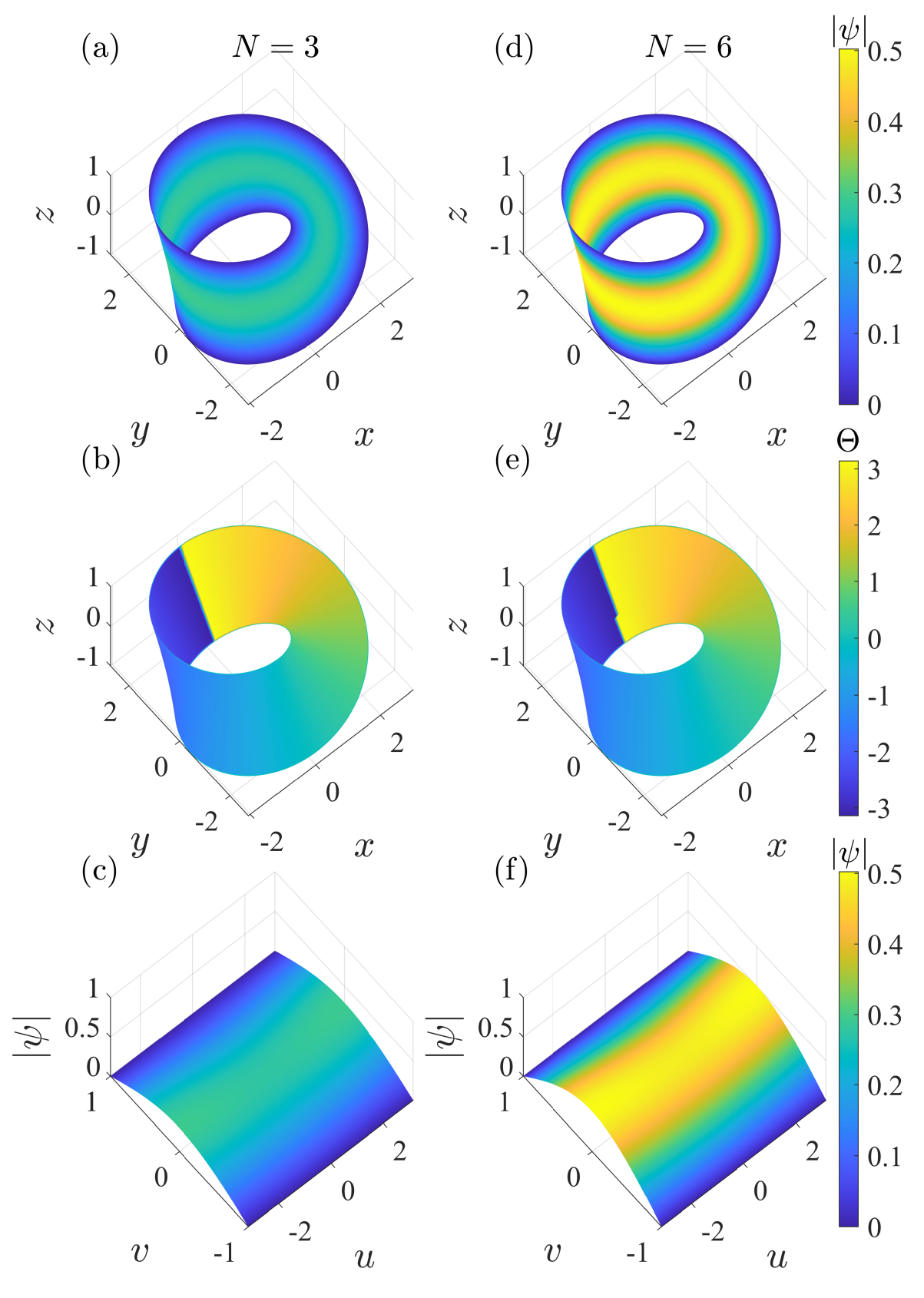}
\caption{Vortex soliton solutions of the full nonlinear equation (\protect
\ref{stationary}), corresponding to quantum numbers ($n=1,m=1$), with norms
and chemical potential (a-c) $N=3,~\protect\mu =1.31$, (d-f) $N=6,~\protect%
\mu =1.19$.}
\label{figure8}
\end{figure}

\begin{figure}[tbp]
\centering
\includegraphics[width=3.4in]{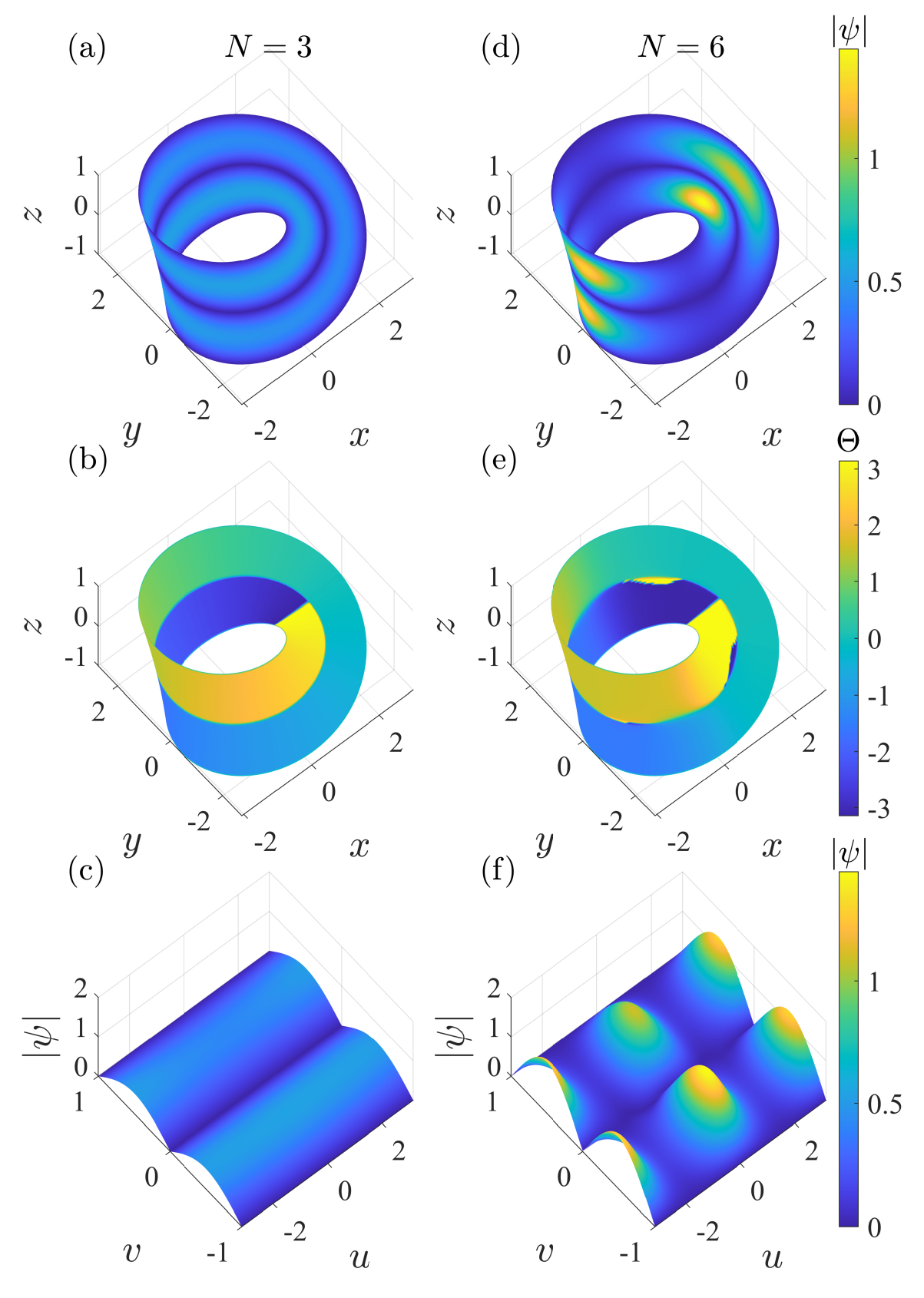}
\caption{Vortex soliton solutions of the full nonlinear equation (\protect
\ref{stationary}), corresponding to quantum numbers ($n=2,m=1/2$), with
norms and chemical potential (a-c) $N=3,~\protect\mu =4.79$, (d-f) $N=6,~%
\protect\mu =4.33$.}
\label{figure9}
\end{figure}

\begin{figure}[tbp]
\centering
\includegraphics[width=3.4in]{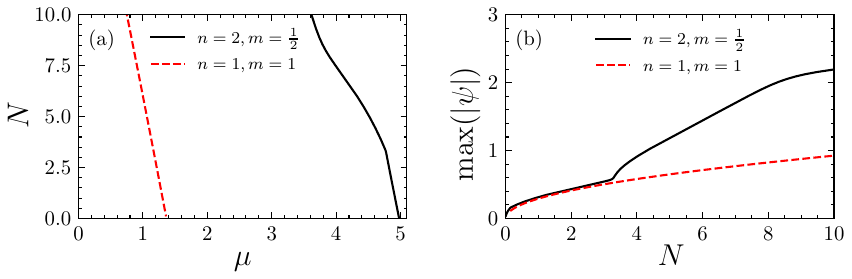}
\caption{(a) The norm of the numerically found vortex solitons, with quantum
numbers $n=1,m=1$ and $n=2,m=1/2$, vs. the chemical potential $\protect\mu $%
. (b) The amplitude $\max (|\protect\psi |)$ of the numerically found vortex
solitons, with quantum numbers $n=1$, $m=1$ and $n=2,m=1/2$, vs. the norm.}
\label{figure10}
\end{figure}

\section{Vortex soliton solutions in the nonlinear system}

In the previous section, we discussed the properties of GS solitons,
including both numerical and variational results. In this section, we
discuss the properties of vortex soliton solutions carried by the MS. Here,
we focus on the vortex states with $n=1,m=1$ and $n=2,m=1/2$, as they
correspond to the lowest-energy vortex states with integer and half-integer
WNs, respectively.

We here again employ the Newton's algorithm within the FEM framework to
produce vortex solitons in the nonlinear system. We choose the linear vortex
states as the initial guess for our Newton iteration. The initial guess for $%
n=1,m=1$ and $n=2,m=1/2$ is $\psi _{1,1}$ and $\psi _{2,1/2}$, respectively.
Additionally, the corresponding initial chemical potentials are taken as the
linear eigenvalues~\eqref{spectrum}, \textit{viz}., $\mu _{1,1}=1.36$ and $%
\mu _{2,1/2}=4.97$, respectively.

By gradually decreasing chemical potential $\mu $, the numerically found
vortex solitons with $n=1,m=1$ and $n=2,m=1/2$ for different norms $N$ are
plotted in Figs.~\ref{figure8} and~\ref{figure9}, respectively. As shown in
Fig.~\ref{figure8}, the vortex solitons with $n=1,m=1$ closely resemble the
linear vortex state, with the wave function being nearly uniformly
distributed in the $u$-direction. Note that, as norm $N$ increases, the wave
functions scale proportionally.

The situation changes for the vortex solitons with $n=2,m=1/2$. As shown in
Figs.~\ref{figure9}, the wave function is uniformly distributed, as above,
in the $u$-direction for $N=3$. However, for $N=6$, the wave function
becomes localized, forming four peaks primarily located along the
cross-section $y=0$, or, equivalently, at $u=0$ and $u=\pm \pi $. Thus,
there is a phase transition from the uniform state to the localized one.

The dependences $N(\mu )$ and $\max (|\psi |)$ on $N$ for vortex solitons
with $n=1,m=1$ and $n=2,m=1/2$ are shown in Fig. \ref{figure10}. The fact
that the plot of $N(\mu )$ in Fig.~\ref{figure10}(a) obeys the VK\
criterion, $dN/d\mu <0$, implies that the vortex solitons also satisfy the
corresponding necessary stability condition. Nevertheless, the full
stability of these modes needs to be confirmed. As $\mu $ decreases, the
norm of the vortex solitons gradually increases. Unlike the GS solitons, the
norm of the vortex solitons has no upper limit, and they do not undergo
collapse. In Fig.~\ref{figure10}(b), the curves of amplitude $\max (|\psi |)$
for the vortex solitons with $n=1,m=1$ and $n=2,m=1/2$ almost overlap at $%
N<3.3$. However, at $N\geq 3.3$, the two curves diverge, with the one which
corresponds to $n=2,m=1/2$ increasing at a faster rate. This fact indicates
that $N=3.3$ is the critical point at which the vortex soliton with $%
n=2,m=1/2$ transitions from the uniform state to the localized one.

\section{The stability analysis for the solitons}

\begin{figure}[bp]
    \centering
    \includegraphics[width=3.4in]{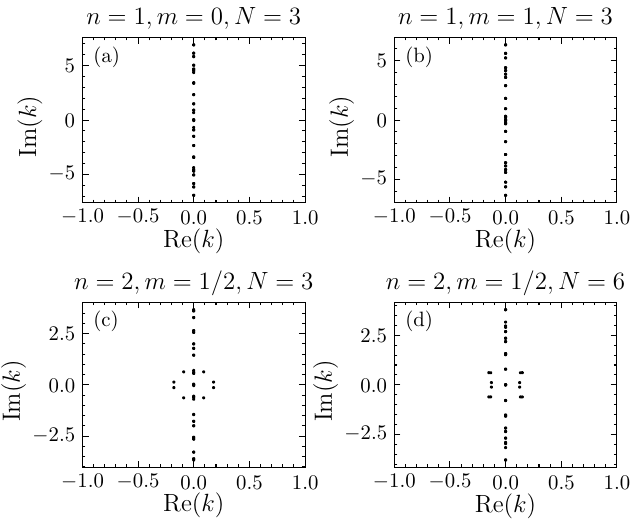}
    \caption{The linear-stability spectra of the GS solitons with (a) $%
    n=1,m=0,N=3$, (b) $n=1,m=1,N=3$, (c) $n=2,m=1/2,N=3$ and (d) $n=2,m=1/2,N=6$%
    . }
    \label{figure11}
\end{figure}

To develop the linear-stability analysis for the real GS solitons, the
perturbed ones are taken as
\begin{equation}
\Psi (u,v,t)=e^{-i\mu t}\left[ \psi (u,v)+\alpha (u,v)e^{kt}+\beta ^{\ast
}(u,v)e^{k^{\ast }t}\right] ,  \label{perturbed}
\end{equation}%
where $\alpha $ and $\beta $ are small perturbations, and $k$ is the
respective (complex) eigenvalue [the stability implying that all eigenvalues
have $\mathrm{Re}(k)=0$]. Inserting the perturbed solution in Eq. %
\eqref{main} and linearizing, one arrives at the linear eigenvalue problem:
\begin{equation}
\left(
\begin{matrix}
\hat{L} & i\psi ^{2} \\
-i\psi ^{\ast 2} & \hat{L}^{\ast }%
\end{matrix}%
\right) \left(
\begin{matrix}
\alpha \\
\beta%
\end{matrix}%
\right) =k\left(
\begin{matrix}
\alpha \\
\beta%
\end{matrix}%
\right) ,  \label{eigenequ}
\end{equation}%
\begin{equation}
\hat{L}\equiv i\mu +\frac{i}{2\lambda }\partial _{u}\left( \frac{1}{\lambda }%
\partial _{u}\right) +\frac{i}{2w^{2}\lambda }\partial _{v}\left( \lambda
\partial _{v}\right) +2i|\psi |^{2}.
\end{equation}%
As $\alpha (u,v)$ and $\beta (u,v)$ also need to satisfy MS\ BC~\eqref{BC},
the eigenvalue problem~\eqref{perturbed} was solved by dint of the FEM
technique.

Figure~\ref{figure11} presents spectra of the stability eigenvalues for (a)
the GS soliton with $n=1,m=0,N=3$, (b) the vortex soliton with $n=1,m=1,N=3$%
, (c) the nearly uniform vortex mode with $n=2,m=1/2,N=3$ and (d) the
well-localized vortex mode with $n=2,m=1/2,N=6$. The respective stationary
states are displayed above in Figs.~\ref{figure3}, \ref{figure8}, and \ref%
{figure9}. The conclusion is that the entire family of the GS solitons is
stable, as well as the family of the vortex solitons with quantum numbers $%
n=1,m=1$. On the other hand, the modes with $n=2,m=1/2$ are fully unstable.

\section{Moving solitons and their collisions}

The defining characteristics of solitons include their remarkable ability to
maintain their shape in the state of motion and collide (quasi-)elastically.
Here, we address the motion and collisions of the solitons in the present
model.

We consider a soliton moving along the circumferential ($u$) direction of
the MS with velocity $c$. To this end, we introduce the Galilean transform
for the coordinates and derivatives:
\begin{equation}
\xi \equiv u-ct,\quad \tau \equiv t,  \label{xi}
\end{equation}
\begin{equation}
\partial _{t}=\partial _{\tau }-c\partial _{\xi },\quad \partial
_{u}=\partial _{\xi }.  \label{pt}
\end{equation}%
casting Eq.~\eqref{main} into the equation for the moving soliton:
\begin{equation}
i\partial _{\tau }\Psi =ic\partial _{\xi }\Psi -\frac{1}{2\lambda }\partial
_{\xi }\left( \frac{1}{\lambda }\partial _{\xi }\Psi \right) -\frac{1}{%
2w^{2}\lambda }\partial _{v}\left( \lambda \partial _{v}\Psi \right) -|\Psi
|^{2}\Psi .  \label{Galilean}
\end{equation}

\begin{figure}[tbp]
\centering
\includegraphics[width=3.4in]{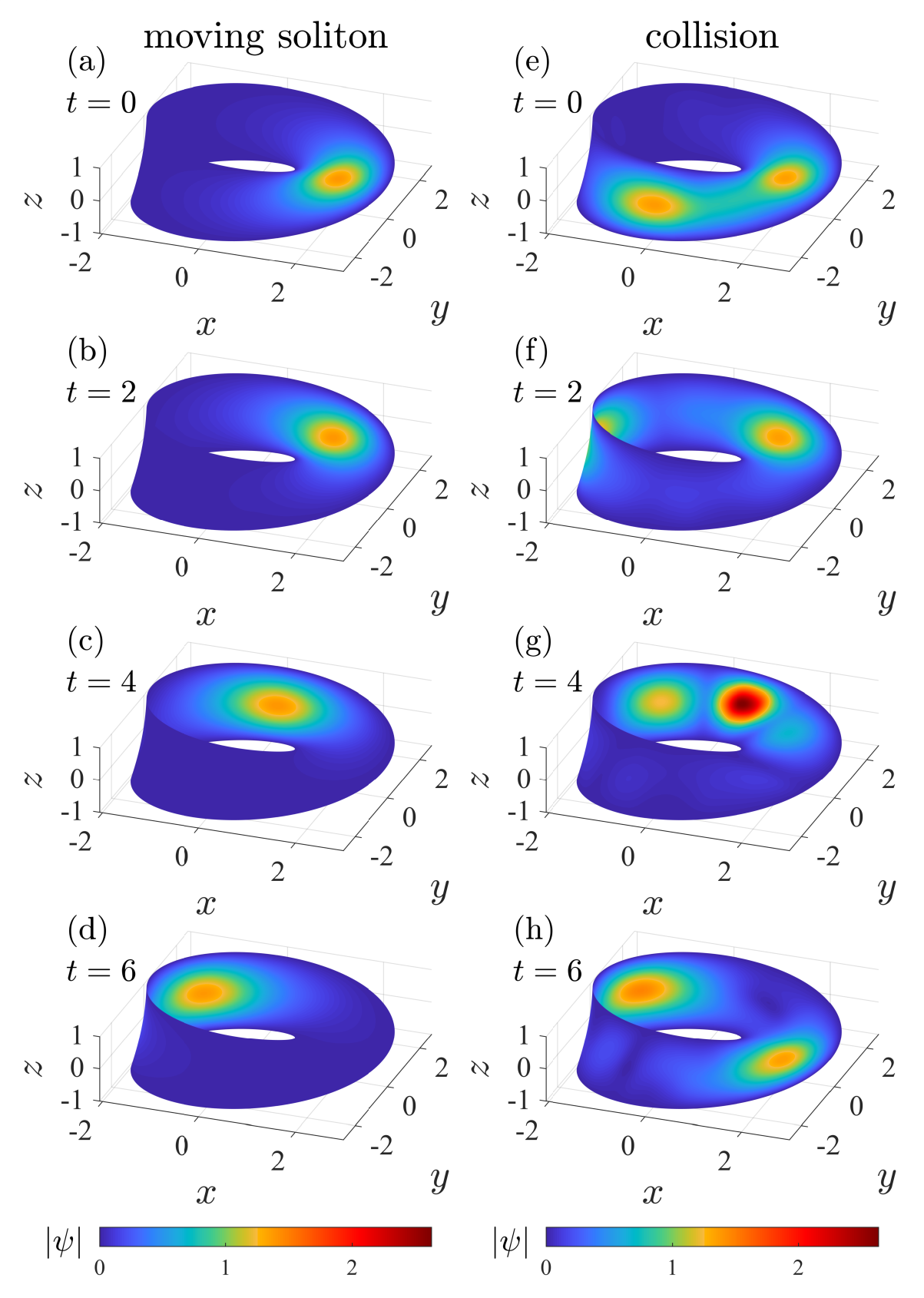}
\caption{The evolution of the density distribution for a single moving
soliton (a-d) and the collision of two solitons (e-h). In (a), the soliton
has velocity $c=\protect\pi /8$, particle number $N=3$, being initially
located at $x=2,y=0,z=0$. In (e), the second soliton is added, with velocity
$c=-\protect\pi /4$, particle number $N=3$, and initial position $%
x=0,y=-2,z=0$}
\label{figure12}
\end{figure}

Using Eq.~\eqref{Galilean} and fixing the total particle number $N$ fixed,
we obtain the respective ground-state soliton $\Psi (\xi ,v,\tau )$ by means
of the imaginary-time evolution method. Then, we use it as the input in the
original reference frame, i.e., in Eq.~\eqref{main}, which produces the
ground-state soliton moving with velocity $c$ along the $u$-direction, as
shown in Fig.~\ref{figure12}(a-d). Here, we set $N=3$ and $c=\pi /8$, with
the respective soliton moving counterclockwise along the MS, with the period
of $T=2\pi /c=16$. At $t=0$, the soliton's center is located at $x=2,y=0,z=0$%
, as shown in Fig.~\ref{figure12}(a). After a quarter-period ($t=4$), it is
positioned at $x=0,y=2,z=0$, as shown in Fig.~\ref{figure12}(c). In the
course of the motion, the shape of the soliton remains unchanged.

Next, we consider collisions of two solitons. Using the method outlined
above, we prepared another soliton with velocity $c=-\pi /4$, particle
number $N=3$, and initial position $x=0,y=-2,z=0$. It moves in the clockwise
direction, reaching the position of $x=0,y=2,z=0$ at the same time, $t=4$,
thus initiating the collide. By superimposing the two solitons and evolving
them according to Eq.~\eqref{main}, the collision of the two solitons is
displayed in Fig.~\ref{figure12}(e-h). As seen in Fig.~\ref{figure12}(g),
the densities of the two solitons do not directly overlap during the
collision. Instead, they form a pattern resembling a standing wave, due to
the opposite momenta of the two solitons. Furthermore, as seen in Fig.~\ref%
{figure12}(h), the shapes of the two solitons remain unchanged after the
collision, and they continue to move forward at their original velocities.
Thus, the collision is quasi-elastic, in spite of the fact that the underlying
equation (\ref{main}) is not integrable.

\section{Conclusion}

In this work, we have introduced the 2D GP (Gross-Pitaevskii) equation with
the self-attractive nonlinearity on the curved surface in the form of the MS
(M\"{o}bius strip), as the model for an atomic BEC trapped on such a
surface. In the linear limit (the solution for which was actually reported
earlier in Ref. \cite{li2012quantum}), the vortex states are characterized
by quantum numbers, $n$ and $m$, corresponding to the transverse and
circumferential directions. The former one, $n$, is always integer, while
the WN (winding number), $m$, is integer for odd $n$ and \emph{half-integer}
for even $n$ (implying the necessity to perform two round trips along MS to
identify the vortical phase). The numerical solution and VA (variational
approximation) were used to produce GS (ground-state) soliton solutions of
the nonlinear GP equation with self-attraction on the MS. The VA predicts 2D
solitons with good accuracy. The attractive cubic nonlinearity leads to the
self-compression of the GS solitons and termination of their existence at
the critical value of the norm, which coincides with that for the TSs
(Townes solitons) in the 2D GP (alias nonlinear Schr\"{o}dinger) equation in
the flat 2D space. However, on the contrary to the unstable TS family in the
flat space, the MS geometry lifts the degeneracy of the soliton family and
makes them \emph{completely stable} (hence, appropriate for the experimental
work), as suggested by the VK (Vakhitov-Kolokolov) criterion and numerical
solution of the corresponding linear-stability problem. The vortex solitons
with quantum numbers $n=1,m=1$ are stable too, while ones with $n=2,m=1/2$
are unstable.

As an extension of the present work, it is possible to simulate the onset
and development of the collapse for the norm exceeding the critical value.
Because the collapse leads to catastrophic self-compression of the wave
function, it may be expected that it will develop similarly to the known
scenario for the nonlinear Schr\"{o}dinger equation in the flat 2D space
\cite{Berge,Sulem,Fibich}. It may also be interesting to construct
delocalized states, including ones with embedded vorticity, in the framework
of the MS GP equation with the repulsive cubic term, while for the
attractive one a remaining relevant problem is the modulational instability
of delocalized states.

\section*{Acknowledgments}

This work was supported by the GuangDong Basic and Applied Basic Research
Foundation through grant No. 2023A1515110198, National Natural Science
Foundation of China through grants Nos. 12274077, 62405054, 12475014,
12075193, 12475026 and 11905032, Natural Science Foundation of Guangdong
Province through grants Nos. 2024A1515030131 and 2021A1515010214, Foundation
for Distinguished Young Talents in Higher Education of Guangdong No.
2024KQNCX150, the Research Found of the Guangdong-Hong Kong-Macao Joint
Laboratory for Intelligent Micro-Nano Optoelectronic Technology through
grant No. 2020B1212030010, and Israel Science Foundation through Grant No.
1695/22.


\appendix

\section{Solving the linear eigenstates by dint of the FEM (finite element
method)}

To apply FEM, we need to define the action $S=\int_{-1}^{+1}\int_{-\pi
}^{+\pi }\mathcal{L}dudv$ corresponding to the linearized equation~\eqref{stationary},
with the Lagrangian
\begin{equation}
\mathcal{L}=\frac{w}{2\lambda }|\partial _{u}\psi |^{2}+\frac{\lambda }{2w}|\partial
_{v}\psi |^{2}-w\lambda \mu |\psi |^{2}.  \label{L}
\end{equation}%
As the $\left\{ u,v\right\} $ space is a rectangle, we perform its uniform
division into the set of $N_{u}\times N_{v}$ small rectangles (\textit{%
elements}), with intervals in the $u$ and $v$-directions being $h_{u}=2\pi
/N_{u}$ and $h_{v}=2/N_{v}$, respectively. Then, the total action is
expressed as the sum of actions over the elements, i.e.,
\begin{equation}
S=\sum_{i,j=1}^{N_{u},N_{v}}S_{ij},\quad
S_{ij}=\int_{v_{j}}^{v_{j+1}}\int_{u_{i}}^{u_{i+1}}\mathcal{L}dudv,  \label{Sij}
\end{equation}%
where the nodes are
\begin{equation}
\left\{
\begin{array}{l}
u_{i}=-\pi +h_{u}(i-1),\quad i=1,2,\cdots ,N_{u}+1, \\
v_{j}=-1+h_{v}(j-1),\quad j=1,2,\cdots ,N_{v}+1.%
\end{array}%
\right.   \label{node}
\end{equation}%
The wave function at the nodes is defined as $\psi (u_{i},v_{i})=\psi _{i,j}$%
, with the continuous wave function $\psi (u,v)$ in intervals $u_{i}\leq
u\leq u_{i+1}$ and $v_{j}\leq v\leq v_{j+1}$ produced by the linear
interpolation:
\begin{equation}
\begin{aligned} &\psi(u,v)\\ &= \frac{(u_{i+1}-u)(v_{j+1}-v)}{h_u h_v}
\psi_{i,j} + \frac{(u-u_i)(v_{j+1}-v)}{h_u h_v} \psi_{i+1,j} \\ & +
\frac{(u_{i+1}-u)(v-v_j)}{h_u h_v} \psi_{i,j+1} + \frac{(u-u_i)(v-v_j)}{h_u
h_v} \psi_{i+1,j+1}. \end{aligned}  \label{ppsi}
\end{equation}%
Higher-order interpolation provides higher accuracy, requiring more nodes in
each element. Note that Lagrangian (\ref{L}) includes the first derivatives
of $\psi $ with respect to $u$ and $v$. Using the interpolated wave function~%
\eqref{ppsi}, we obtain
\begin{equation}
\begin{aligned} \partial_u\psi&\!=\! \frac{v_{j+1}-v}{h_u h_v}
(\psi_{i+1,j}\!-\!\psi_{i,j}) \!+\! \frac{v-v_j}{h_u
h_v}(\psi_{i+1,j+1}\!-\!\psi_{i,j+1}),\\ \partial_v\psi&\!=\!
\frac{u_{i+1}-u}{h_u h_v} (\psi_{i,j+1}\!-\!\psi_{i,j}) \!+\!
\frac{u-u_i}{h_u h_v}(\psi_{i+1,j+1}\!-\!\psi_{i+1,j}).\\ \end{aligned}
\label{dpsi}
\end{equation}

Now, substituting the wave function~\eqref{ppsi} and its derivatives~%
\eqref{dpsi} in Eqs.~\eqref{L} and~\eqref{Sij}, and integrating, we obtain:
\begin{equation}
S_{ij}=%
\begin{pmatrix}
\psi_{i,j}^{\ast } \\
\psi_{i,j+1}^{\ast } \\
\psi_{i+1,j}^{\ast } \\
\psi_{i+1,j+1}^{\ast }%
\end{pmatrix}%
^{T}\left( H^{ij}-\mu U^{ij}\right)
\begin{pmatrix}
\psi _{i,j} \\
\psi _{i,j+1} \\
\psi _{i+1,j} \\
\psi _{i+1,j+1}%
\end{pmatrix}%
,
\end{equation}%
where the local matrices $H^{ij}$ and $U^{ij}$ can be expressed as 
\begin{equation}
\begin{aligned}
H^{ij} &= \displaystyle\int_{v_{j}}^{v_{j+1}} \displaystyle\int_{u_{i}}^{u_{i+1}} \left\{ 
\frac{w}{2\lambda h_u^2 h_v^2} 
\begin{pmatrix}
v - v_{j+1} \\
v_j-v \\
v_{j+1} - v \\
v - v_j
\end{pmatrix}
\begin{pmatrix}
    v - v_{j+1} \\
    v_j-v \\
    v_{j+1} - v \\
    v - v_j
\end{pmatrix}^T \right. \\
&\quad + \left. \frac{\lambda}{2w h_u^2 h_v^2} 
\begin{pmatrix}
u - u_{i+1} \\
u_{i+1} - u \\
u_i - u \\
u - u_i
\end{pmatrix}
\begin{pmatrix}
u - u_{i+1} \\
u_{i+1} - u \\
u_i - u \\
u - u_i
\end{pmatrix}^T 
\right\} du \, dv,
\end{aligned}
\end{equation}
and
\begin{equation}
    \begin{aligned}
   & U^{ij} = \int_{v_j}^{v_{j+1}} \int_{u_i}^{u_{i+1}} 
    \frac{w\lambda \, du, dv}{h_u^2 h_v^2} 
   \\
    & \times
    \begin{bmatrix}
        (u_{i+1} - u)(v_{j+1} - v) \\
        (u_{i+1} - u)(v - v_j) \\
        (u - u_i)(v_{j+1} - v) \\
        (u - u_i)(v - v_j)
        \end{bmatrix} 
        \begin{bmatrix}
    (u_{i+1} - u)(v_{j+1} - v) \\
    (u_{i+1} - u)(v - v_j) \\
    (u - u_i)(v_{j+1} - v) \\
    (u - u_i)(v - v_j)
    \end{bmatrix}^T.
    \end{aligned}
\end{equation}

At this point, we need to incorporate the boundary conditions on the M\"{o}%
bius strip~\eqref{BC}, which, in terms of the finite-element framework, is:
\begin{equation}
\psi _{i,1}=\psi _{i,N_{v}+1}=0,\quad \psi _{N_{u}+1,j}=\psi _{1,N_{v}+2-j}.
\end{equation}%
We need to define $\{\psi _{i,j}\}$ as a column matrix, with indices ranging
over $i=1,2,\dots ,N_{u}$ and $j=2,\dots ,N_{v}$ (here, the boundary points
are disregarded). In this way, the total action is expressed as
\begin{equation}
S=\{\psi _{i,j}^{\ast }\}^{T}\left( H-\mu U\right) \{\psi _{i,j}\},
\end{equation}%
where the global matrices $H$ and $U$ are assembled by the local matrices 
$H^{ij}$ and $U^{ij}$~\cite{fem_QM}. We now invoke the principle of stationary action and
vary the action with respect to the nodal values of wavefunction $\psi
_{i,j}^{\ast }$., which leads to the generalized eigenvalue equation,
\begin{equation}
H\{\psi _{i,j}\}=\mu U\{\psi _{i,j}\}.
\end{equation}%
By numerically solving this equation, we obtain the results for the
eigenvalues and eigenstates, which are shown in Fig.~\ref{figure1}.


\end{document}